\documentclass[prd,twocolumn,showpacs,nofootinbib]{revtex4}

\usepackage{amsfonts}
\usepackage{amsmath}
\usepackage{amssymb}
\usepackage{upgreek}
\usepackage{bm}
\usepackage{dcolumn}
\usepackage{epsfig}
\usepackage{graphicx}
\usepackage{graphics}
\newcommand{\mb}[1]{\mathbf{#1}}

\begin{document}

\title{Towards a unified treatment of gravitational-wave data analysis}

\author{Neil J.~Cornish}
\affiliation{Department of Physics, Montana State University, 
Bozeman, MT 59718, USA.}
\author{Joseph D.~Romano}
\affiliation{Department of Physics and Astronomy 
and Center for Gravitational-Wave Astronomy,
University of Texas at Brownsville, Brownsville, TX 78520, USA.}

\date{\today}

%%%%%%%%%%%%%%%%%%%%%%%%%%%%%%%%%%%%%%%%%%%%%%%
\begin{abstract}
We present a unified description of gravitational-wave data analysis that unites the template-based analysis used to detect 
deterministic signals from well-modeled sources, such as binary-black-hole mergers, with the cross-correlation analysis
used to detect stochastic gravitational-wave backgrounds. We also discuss the connection between template-based analyses
and those that target poorly-modeled bursts of gravitational waves, and suggest a new approach for detecting burst signals.
\end{abstract}
%%%%%%%%%%%%%%%%%%%%%%%%%%%%%%%%%%%%%%%%%%%%%%%

\maketitle

%%%%%%%%%%%%%%%%%%%%%%%%%%%%%%%%%%%%%%%%%%%%%%%

Gravitational-wave data analysis is conventionally divided into 
three classes that depend on the nature of the signal: 
(i) well-modeled deterministic signals, such as those
from compact binary inspirals; (ii) poorly-modeled deterministic 
signals, such as those from core-collapse supernovae; 
and (iii) stochastic signals, such as those from a phase
transition in the early Universe. 
Here we will argue that this division is rather artificial 
and unnecessary, and suggest that a unified treatment 
can yield deeper insights.
The elements needed to unify cases (i) and (ii) can be found in 
Refs.~\cite{Flanagan:1997kp,Klimenko:2005xv,searle,searle2}.
Here we provide a unification of cases (i) and (iii) 
using hierarchical Bayesian modeling~\cite{Adams:2012qw}.

The motivation for developing a unified description of gravitational-wave data analysis is two-fold.
First there is the pedagogical value of a coherent picture that emphasizes the common
foundation of the disparate analysis techniques found in the literature, and second, the unified picture
can provide a deeper understanding that may suggest new approaches. To illustrate the latter point
we conclude our discussion by proposing a novel technique for detecting un-modeled ``bursts'' of gravitational waves.

In the conventional picture, signals for which we have waveform 
templates are analyzed using a matched-filter statistic~\cite{kip},
un-modeled signals are characterized in terms of an excess power statistic~\cite{Anderson:2000yy},
and stochastic signals are analyzed using a cross-correlation statistic between pairs of detectors~\cite{Michelson}. 
The connection between the various forms of analysis is not 
immediately apparent, especially when described in a frequentist framework.

In the case of un-modeled signals, the analyses usually focus on short duration ``bursts''
of gravitational-wave energy that are localized in a time-frequency representation of the data.
The connection between a waveform template-based search and a burst search becomes
apparent in the case of fully-coherent network analyses, where it becomes possible to
solve for the gravitational-wave signal by either maximizing the likelihood~\cite{Klimenko:2005xv}
or locating regions of high posterior density~\cite{searle,searle2}. To obtain meaningful results
these analyses require constraints or priors on the signal models. The waveform template-based
analyses are recovered in the limit that the signal priors become highly informative, ultimately
mapping the individual signal samples $h_i$ to a small number of physical parameters $\vec{\lambda}$
that describe the signal: $h_i(\vec{\lambda})$.

To develop the connection between the cases (i) and (iii) we found it advantageous to adopt a
hierarchical Bayesian analysis framework~\cite{Good:1965,Morris:1992} with parametrized 
likelihood functions and a parameterized signal prior.  We begin with the simplest case imaginable: that of two co-located 
and co-aligned detectors, which each provide a single datum 
$s_1 = n_1 + h$, $s_2= n_2 + h$, respectively.
The noise in the detectors is assumed to be Gaussian random and 
independent, with zero mean and variance $\sigma^2_i$. 
The signal, $h$, which is common to both detectors, 
is also assumed to come from a Gaussian distribution with 
zero mean and variance $\sigma^2_h$.
The quantities $\sigma_i, \sigma_h$
are our model hyper-parameters, which are to be determined from 
the data. 

Now suppose that we adopt a waveform template $h$ to describe 
the gravitational wave
signal, and form the residuals $r_i = s_i -h$. 
We demand that the residuals be consistent with the 
probability distribution for the noise~\cite{Finn:1992wt}, 
which in this case results in a multi-variate Gaussian 
likelihood function:
\begin{equation}\label{temp}
p(s\vert \sigma_i, h) = \frac{1}{\sqrt{(2\pi)^2\, {\rm det}{C'}} }\, 
e^{-\frac{1}{2}\, r_i (C'^{-1})_{ij} r_j},
\end{equation}
with
\begin{equation}
C'_{ij} = \delta_{ij} \sigma_i^2 .
\end{equation}
Since our template is stochastic, we are not interested in 
the particular value of $h$, but rather, in the overall amplitude 
$\sigma_h$. 
Thus, using the parameterized signal prior
\begin{equation}\label{sprior}
p(h\vert \sigma_h)
=\frac{1}{\sqrt{2\pi\sigma_h^2}} e^{-h^2/2\sigma_h^2}
\end{equation}
we marginalize over $h$:
\begin{eqnarray}\label{marg}
p(s|\sigma_i,\sigma_h)
&\equiv&
\int p(s\vert \sigma_i, h)\, p(h\vert \sigma_h)\, dh\nonumber \\
&=&\frac{1}{2\pi \,\sigma_1\sigma_2}
\int e^{-(s_1-h)^2/2\sigma_1^2}e^{-(s_2-h)^2/2\sigma_2^2}\nonumber \\
&&\times \frac{1}{\sqrt{2\pi}\,\sigma_h}  e^{-h^2/2\sigma_h^2}\, dh \nonumber \\
&=&\frac{1}{\sqrt{(2\pi)^2\, {\rm det}C} }\, e^{-\frac{1}{2}\, s_i (C^{-1})_{ij} s_j},
\label{corr}
\end{eqnarray}
where the matrix $C$ has components
\begin{equation}
C_{ij} = \delta_{ij} \sigma_i^2 +\sigma_h^2\, .
\end{equation}
The likelihood function (\ref{corr}) has the standard form used in 
cross-correlation analyses for stochastic 
signals~\cite{vanHaasteren:2008yh,Adams:2010vc}.
Thus we see that a template-based analysis (\ref{temp}) using 
stochastic templates is equivalent to a cross-correlation analysis 
(\ref{corr}) without templates.

The generalization to $N$ co-aligned and co-located
detectors is straightforward. 
Expression (\ref{temp}) becomes
\begin{equation}\label{temp2}
p(s\vert \sigma_i, h) =
\frac{1}{\sqrt{ (2\pi)^N{\rm det}C'} }\, e^{-\frac{1}{2}\, r_i (C'{}^{-1})_{ij} r_j},
\end{equation}
and the marginalization proceeds as follows:
\begin{eqnarray}
\label{marg2}
&p(s\vert \sigma_i,\sigma_h)&=\int  p(s\vert \sigma_i,h)\, 
p(h\vert \sigma_h)\, dh\nonumber \\
&&= \frac{e^{-\frac{1}{2}\,s_i (C^{-1})_{ij} s_j}}{\sqrt{(2\pi)^N {\rm det} C'} } 
\int \frac{dh}{\sqrt{2\pi}\,\sigma_h} \nonumber \\
&&\!\!\!\!\!\!\!\!\!\!\!
\times \exp\left(-\frac{{\rm det} C}{2 \sigma_h^2 {\rm det}C' } 
\left( h- \frac{\sigma_h^2{\rm det}C' }{{\rm det} C}
\sum_i \frac{s_i}{\sigma_i^2}\right)^2 \right)    \nonumber \\
&& \nonumber \\
&&=  \frac{ e^{-\frac{1}{2}\,s_i (C^{-1})_{ij} s_j}}{\sqrt{(2\pi)^N{\rm det}C} }\, 
\int \frac{1}{\sqrt{2\pi}\,\sigma_h} e^{-h'^2/2\sigma_h^2}     \, dh' \nonumber \\
&&= \frac{1}{\sqrt{(2\pi)^N {\rm det}C} }\, e^{-\frac{1}{2}\, s_i (C^{-1})_{ij} s_j}\,.
\label{corr2}
\end{eqnarray}
The marginalization involves a completion of the square, 
followed by a change of variables from $h$ to $h'$ via a shift, 
then a rescaling. 
It is the rescaling that takes $\sqrt{{\rm det} C'}$ to $\sqrt{{\rm det} C}$ 
in the denominator. 

The generalization to multiple data points is trivial.
The individual variances 
$\sigma_i^2$, $\sigma_h^2$ 
are replaced by variance-covariance {\em matrices}
$\boldsymbol{\Sigma}_i$, $\boldsymbol{\Sigma}_h$,
with corresponding replacements for $C'$ and $C$:
\begin{equation}
\mathbf{C'}_{ij} = \delta_{ij}\,\boldsymbol{\Sigma}_i\,,
\quad
\mathbf{C}_{ij} = \delta_{ij}\,\boldsymbol{\Sigma}_i
+\boldsymbol{\Sigma}_h\,.
\end{equation}
Assuming time stationarity,
these variance-covariance matrices can be written
in the frequency domain as one-sided 
spectral density functions 
\begin{align}
&C'_{ij}(f) = \delta_{ij} S_{n_i}(f)\,,
\\
&C_{ij}(f) = \delta_{ij} S_{n_i}(f) +S_h(f)\,.
\end{align}
In terms of these functions the likelihood 
and marginalized likelihood are given by
\begin{align}
p(s\vert \sigma_i, h)&= \prod_{f}  
\frac{1}{(2\pi)^N\,{\rm det}C'(f) }\, 
e^{-2 \, \tilde{r}_i(f) C'_{ij}(f)^{-1} \tilde{r}_j^*(f)}\,,
\label{temp2f}
\\
p(s\vert \sigma_i, \sigma_h)&= \prod_{f}  
\frac{1}{(2\pi)^N\,{\rm det}C(f) }\, 
e^{-2 \, \tilde{s}_i(f) C_{ij}(f)^{-1} \tilde{s}_j^*(f)}\,.
\label{corr2f}
\end{align}

The extension to multiple polarization states and 
signals with a spatial distribution is also relatively straightforward.
To simplify the discussion, let us start by considering
co-located, but no longer co-aligned detectors, 
and write 
\begin{equation}
h_i = F_i^+ h_+ + F_i^\times h_\times\,,
\end{equation}
where $F_i^+$ and $F_i^\times$ are the antenna beam patterns for
the two polarization states for detector $i$ 
in the long-wavelength limit. 
Then the signal prior for $h\equiv(h_+,h_\times)$ 
takes the form
\begin{equation}\label{margpc}
p(h\vert \sigma_h) =
\frac{1}{2\pi \sigma^2_h} 
e^{-(h_+^2+h_\times^2)/2\sigma_h^2} \,.
\end{equation}
Marginalization over $h$ 
now involves integration over both $h_+$ and $h_\times$.
For example, we start with the residual 
$r_i = s_i - F_i^+ h_+ - F_i^\times h_\times$ and a 
diagonal correlation matrix $C'_{ij}=\delta_{ij}\sigma^2_i$.
After doing the integral over $h_\times$ we have a 
{\em new} residual $r_i' = s_i - F_i^+ h_+$, 
and the correlation matrix picks up the off-diagonal 
term $F_i^\times F_j^\times \sigma_h^2$. 
Then the $h_+$ integration takes us to 
$r_i^{''} = s_i$, and the correlation matrix 
\begin{equation}
C_{ij} = \delta_{ij} \sigma_i^2 +(F^+_i F^+_j +F^\times_i F^\times_j)\sigma_h^2\, .
\end{equation}
To handle the case of sources that are distributed 
across the sky, we first write 
\begin{equation}
h_i =\sum_{n=1}^M  F_i^+(\hat{n}) h^n_+ + F_i^\times({\hat n}) h^n_\times\,,
\end{equation}
where $n$ labels the sky location in the direction $\hat{n}$. 
The continuum limit can be found by taking $M \rightarrow \infty$ 
and replacing the sum by an integral. 
For an isotropic background the signal prior for 
$h\equiv (h_+^n,h_\times^n)$ is given by
\begin{equation}
\label{margpcs}
p(h\vert \sigma_h)=\prod_{n=1}^M  \frac{1}{2\pi \sigma^2_h}  
e^{-((h^n_+)^2+(h^n_\times)^2)/2\sigma_h^2}\,.
\end{equation}
Repeated application of the marginalization as before
yields the likelihood (\ref{corr2}) with the correlation matrix
\begin{equation}\label{ovred}
C_{ij} = \delta_{ij} \sigma_i^2 +
\frac{\bar\sigma_h^2}{4\pi} \int \, d\Omega_{\hat n}
(F^+_{i}(\hat{n}) F^+_{j}(\hat{n})  +F^\times_i(\hat{n})  F^\times_j(\hat{n}) )\,,
\end{equation}
where $\bar\sigma_h^2=\sigma_h^2\,M$.
We have taken the continuum limit in writing the final 
expression for the correlation matrix, noting that as
$M\rightarrow\infty$, the variance $\bar\sigma_h^2$ is 
held constant.
The diagonal terms in the signal-dependent part of 
(\ref{ovred}) are the sky (and polarization) averaged 
response functions for the detectors. 
The off-diagonal term in (\ref{ovred}) is proportional
to the {\em overlap reduction function} for co-located, 
but mis-aligned detectors.
The derivation for anisotropic backgrounds is very similar, the only
difference being that the $\sigma_h$ are different for each sky location, 
leading to the correlation matrix
\begin{eqnarray}
&&C_{ij} = \delta_{ij} \sigma_i^2 +  
\frac{1}{4\pi} \int \, d\Omega_{\hat n} 
(F^+_{i}(\hat{n}) F^+_{j}(\hat{n})  \nonumber \\
&& \quad \quad +F^\times_i(\hat{n})  F^\times_j(\hat{n}) )\, 
\bar\sigma^2_h(\hat{n})\, .
\end{eqnarray}

Finally, for spatially separated and mis-aligned detectors 
we can adopt a coordinate system where the detectors are 
located at $\vec{x}_1$ and $\vec{x}_2$, so that
in the Fourier domain the signals can be written as
\begin{equation}
\tilde{h}_i(f)=\sum_{n=1}^M  
(F^+_i(\hat{n})  \tilde{h}^n_+(f) + F^\times_i(\hat{n}) \tilde{h}^n_\times(f))
e^{2\pi i f \vec{x}_i\cdot {\hat n}}\, .
\end{equation}
Marginalizing over the stochastic signal for an isotropic
background yields the likelihood (\ref{corr2f}) with 
the Hermitian correlation matrix
\begin{equation}\label{ovredC}
C_{ij}(f) = \delta_{ij} S_{n_i}(f) + \gamma_{ij}(f) S_h(f)\,,
\end{equation}
where $\gamma_{ij}(f)$ is the overlap reduction function~\cite{Christensen:1992wi,Flanagan:1993ix}
\begin{eqnarray}\label{ovredf}
\gamma_{ij}(f) &=& \frac{ 1}{4\pi} \int \,  
(F^+_{i}(\hat{n}) F^+_{j}(\hat{n})  +F^\times_i(\hat{n})  F^\times_j(\hat{n}) ) 
\nonumber \\
&& \quad \quad \times e^{2\pi i f (\vec{x}_i-\vec{x}_j)\cdot {\hat n}}
\, d\Omega_{\hat n}\, .
\end{eqnarray}
If the background were anisotropic, 
the correlation matrix would have the form
\begin{equation}\label{ovredfaC}
C_{ij}(f) = \delta_{ij} S_{n_i}(f) + \kappa_{ij}(f),
\end{equation}
with
\begin{eqnarray}\label{ovredfa}
\kappa_{ij}(f) &=& \frac{ 1}{4\pi} \int \,  
(F^+_{i}(\hat{n}) F^+_{j}(\hat{n})  +F^\times_i(\hat{n})  F^\times_j(\hat{n}) ) 
\nonumber \\
&& \quad \quad \times S_h(\hat{n}, f) \, 
e^{2\pi i f (\vec{x}_i-\vec{x}_j)\cdot {\hat n}}\, d\Omega_{\hat n}\, .
\end{eqnarray}

Note that all of the above results are special 
cases of the general result for the convolution of 
two multi-variate Gaussian distributions:
\begin{align}
&\int 
\frac{e^{-\frac{1}{2}({\mb x}-{\mb F}\cdot{\mb y})^T
\mb{D}^{-1}({\mb x}-{\mb F}\cdot{\mb y})}}
{\sqrt{(2\pi)^{N_x}\det{\mb D}}}
\frac{e^{-\frac{1}{2}{\mb y}^T\mb{E}^{-1}{\mb y}}}
{\sqrt{(2\pi)^{N_y}\det{\mb E}}}
\, d{\mb y}\,
\nonumber\\
&\qquad\qquad\qquad\qquad
=\frac{1}{\sqrt{(2\pi)^{N_x}\det{\mb C}}}
e^{-\frac{1}{2}{\mb x}^T\mb{C}^{-1}{\mb x}}
\end{align}
where 
\begin{equation}
{\mb C}^{-1} = {\mb D}^{-1} 
-{\mb D}^{-1}{\mb F}({\mb E}^{-1}+ {\mb F}^T{\mb D}^{-1}{\mb F})^{-1}
{\mb F}^T{\mb D}^{-1}
\end{equation}
and
\begin{equation}
{\mb C} = {\mb D} + {\mb F}{\mb E}{\mb F}^T\,.
\end{equation}

The connection between a stochastic template-based analysis and a 
cross-correlation analysis was partially developed in 
Refs.~\cite{Drasco:2002yd, Allen:2002jw},
where a connection was made between a template-based maximum-likelihood 
estimator and the cross-correlation statistic used to search for stochastic
backgrounds.
However, the analysis was limited to a pair of co-located and co-aligned 
detectors, and the possibility of a full unification was not developed.

Establishing a unified description of gravitational-wave data analysis 
has value beyond pedagogy. Seeing the connection between the analyses
can suggest new approaches. Of particular promise are novel approaches that
fall between the conventional divisions. In the unified approach, the analysis begins
with a model for the instrument noise, which then becomes the likelihood for the
residuals $r=s-h$. The specification of the prior on the signal model then completes
the model, giving a continuum of analysis techniques that range from the highly
informative waveform priors $p(h\vert \vec{\lambda}) = \delta(h-h(\vec{\lambda}))$ of a standard
matched filter analysis to the stochastic prior $p(h\vert \sigma_h)$ that yields the standard cross-correlation
analysis. While it may be difficult to reverse engineer the form of the signal prior that yields
some of the existing hybrid analysis techniques~\cite{Brady:1998nj, Krishnan:2004sv, Cutler:2005pn, Ballmer:2005uw,
Dhurandhar:2007vb, Thrane:2010ri, Prix:2011qv}, it is likely that such a mapping will always exist as we have almost
infinite freedom in the choice of the signal prior. Specific examples of such reverse engineering of the signal priors can be
found in Ref.~\cite{searle2}. Rather than trying to recover existing approaches, a more promising avenue for future
research is to explore alternative choices for the signal prior that may yield useful analysis techniques.

To give a concrete example of a new analysis approach that is suggested by the unified picture, we propose a simple
signal prior for un-modeled bursts of gravitational wave radiation. We picture these signals as occupying a
relatively small area in time-frequency space, so it is natural to work in a wavelet basis where the signal in
each detector can be written as $s_{\mu ij}$, where the Greek index $\mu$ labels the detector and the Roman
indices $i, j$ denote the location in time and frequency respectively. For a co-located and co-aligned two detector network
the likelihood is then
\begin{equation}\label{likew}
p(s\vert \sigma_{\mu j}, \sigma_{\nu l},h) = \frac{e^{-\frac{1}{2}\, r_{(\mu ij)} (C'^{-1})_{(\mu ij)(\nu k l)}r_{(\nu kl)}}}{\sqrt{(2\pi)^{2N}\, {\rm det}{C'}} }\,  ,
\end{equation}
with
\begin{equation}\label{corw}
 C'_{(\mu ij)(\nu k l)} =  \sigma^2_{\mu j}\, \delta_{\mu\nu} \delta_{ik}\delta_{jl} .
\end{equation}
Here $N$ is the number of wavelet components.
The expression for the correlation matrix is only approximate as there will be a some noise correlation between frequency layers $j,l$, but
in a well-chosen wavelet basis the correlation is negligible. 
For the signal model we assume that the wavelet amplitudes are Gaussian random distributed:
\begin{equation}\label{spriorw}
p(h_{(ij)} \vert {\sigma_h}_{(ij)}(f_c, t_c,\Delta f, \Delta t))
=\frac{1}{\sqrt{2\pi } \,{\sigma_h}_{(ij)}}e^{-h_{(ij)}^2/2{\sigma_h}_{(ij)}^2}
\end{equation}
with amplitudes ${\sigma_h}_{(ij)}$ that depend on a central frequency $f_c$, central time $t_c$, 
and widths $\Delta f$ and $\Delta t$:
\begin{equation}
 {\sigma_h}_{(ij)}(f_c, t_c,\Delta f, \Delta t) = \sigma e^{-((t_i-t_c)^2/2\Delta t^2+(f_j-f_c)^2/2\Delta f^2)} \, .
\end{equation}
Marginalizing over $h$ yields
\begin{equation}\label{likewm}
p(s\vert \sigma_{\mu j}, \sigma_{\nu l},h) = \frac{e^{-\frac{1}{2}\, s_{(\mu ij)} (C^{-1})_{(\mu ij)(\nu k l)}s_{(\nu kl)}}}{\sqrt{(2\pi)^N\, {\rm det}{C}} }\,  ,
\end{equation}
with
\begin{equation}\label{corwm}
 C_{(\mu ij)(\nu k l)} =  (\sigma^2_{\mu j}\, \delta_{\mu\nu} + {\sigma_h}_{(ij)}^2) \delta_{ik}\delta_{jl} \, .
\end{equation}
For frequencies and times $f_j$, $t_i$ that are far from the central values $f_c$, $t_c$ the correlation matrix reduces to that of uncorrelated noise.
In effect we end up with a cross-correlation search that targets a small region in time-frequency. The analysis is easily extended to $M$ mis-aligned and
spatially distributed detectors by linearly transforming the data to form $M-2$ null streams and 2 signal streams~\cite{Klimenko:2005xv}. The cross-correlation
statistic then depends on 8 parameters: the sky location $\theta,\phi$ and polarization angle $\psi$; the central values $f_c,t_c$; the widths $\Delta f, \Delta t$ and
the overall amplitude $\sigma$. Our example burst search is essentially equivalent to applying a radiometer style search~\cite{Ballmer:2005uw} to data that has
been convolved with a Gaussian time-frequency window function, and is also similar to the {\em STAMP} algorithm~\cite{Thrane:2010ri}.

%%%%%%%%%%%%%%%%%%%%%%%%%%%%%
\acknowledgments
NJC acknowledges support from NSF Award PHY-1205993  and NASA grant NNX10AH15G.
JDR acknowledges support from NSF Awards PHY-1205585 and CREST HRD-1242090.
We would also like to thank Nelson Christensen and Eric Thrane for comments
on an earlier draft of this paper.
 
%%%%%%%%%%%%%%%%%%%%%%%%%%%%%
%%%%%%%%%%%%%%%%%%%%%%%%%%%%%

\end{document}